\documentclass[aps, prd, twocolumn, superscriptaddress,tightenlines, preprintnumbers,nofootnoteinbib]{revtex4-1}
\usepackage[T1]{fontenc}
\usepackage{microtype}
\usepackage[textsize=scriptsize,backgroundcolor=red!70,linecolor=red]{todonotes}
\usepackage{booktabs}
\usepackage{dcolumn}
\usepackage{amsmath}
\usepackage{amsfonts}
\usepackage{amssymb}
\usepackage{graphicx}
\usepackage{ulem}
\usepackage{hyperref}
\usepackage[all]{hypcap}
\usepackage{paralist}
\usepackage{multirow}
\usepackage{mathrsfs}
\usepackage{graphicx}
\usepackage{dcolumn}
\usepackage{bm}
\usepackage{epsf}
\usepackage{dcolumn}
\newcommand{\zbl}[1]{\textcolor{black}{#1}}

\begin{document}

\DeclareGraphicsExtensions{.eps,.ps}

\title{Tensor interaction in coherent elastic neutrino-nucleus scattering}

\author{Jiajun Liao}
\email[Contact author: ]{liaojiajun@mail.sysu.edu.cn.}
\affiliation{School of Physics, Sun Yat-sen University, Guangzhou, 510275, China}

\author{Jian Tang}
\email[Contact author: ]{tangjian5@mail.sysu.edu.cn.}
\affiliation{School of Physics, Sun Yat-sen University, Guangzhou, 510275, China}
 
\author{Bing-Long Zhang}
\email[Contact author: ]{zhangblong@mail2.sysu.edu.cn.}
\affiliation{School of Physics, Sun Yat-sen University, Guangzhou, 510275, China}

\begin{abstract}
Neutrino tensor interactions have gained prominence in the study of coherent elastic neutrino-nucleus scattering (CE$\nu$NS) recently. 
We perform a systematical examination of the nuclear effect, which plays a crucial role in evaluating the cross section of CE$\nu$NS in the presence of tensor interactions.
Our analysis reveals that the CE$\nu$NS cross section induced by tensor interactions is not entirely nuclear spin-suppressed and can be enhanced by a few orders of magnitude compared to the conventional studies. 
The neutrino magnetic moment induced by the loop effect of tensor interactions, 
is also taken into account due to its sizable contribution to the CE$\nu$NS cross section. 
We also employ data from the COHERENT experiment and recent observations of solar $^8$B neutrinos from dark matter direct detection experiments to scrutinize the parameter space of neutrino tensor interactions.  
\end{abstract}
\maketitle

{\bf Introduction.} 
The discovery of the neutrino oscillation clearly indicates the existence of new physics beyond the standard model (SM), triggering substantial theoretical and experimental efforts to explore the nature of neutrinos. A model-independent way to explore new physics in the neutrino sector is parameterized in the form of generalized neutrino interactions~\cite{Lee:1956qn,Bergmann:1999rz,Lindner:2016wff,AristizabalSierra:2018eqm}, i.e., four fermion operators with scalar, pseudo-scalar, vector, axial-vector and tensor Lorentz-invariant structures. Among these interactions, the tensor interactions historically received comparatively less attention but attracted numerous interests in the literature in recent years. Neutrino tensor interactions can arise from lepto-quark models ~\cite{Herczeg:2001vk}, left-right symmetric models~\cite{Xu:2019dxe,Chen:2022xkk}, \zbl{SM effective field theory extended with right-handed neutrinos (SMNEFT)~\cite{Bischer:2019ttk, Han:2020pff}}, and unparticle physics models~\cite{Barranco:2011wx}. The parameter space of tensor interactions has been explored by diverse phenomenological channels, including deep inelastic scattering~\cite{Han:2020pff,Escrihuela:2021mud,Escrihuela:2023sfb}, muon decay~\cite{Gagliardi:2005fg, TWIST:2011aa}, neutrino mass~\cite{Ito:2004sh, Prezeau:2004md}, $\beta$ decay~\cite{ Carnoy:1991jd, Adelberger:1999ud,Cirigliano:2013xha, Gonzalez-Alonso:2018omy, Banerjee:2023lrk, Aker:2024dbg}, etc.

\zbl{Recent observations of the coherent elastic neutrino-nucleus scattering process (CE$\nu$NS), arising from the exchange of a vector $Z$ boson, provide a novel probe to new physics in the neutrino sector~\cite{Abdullah:2022zue}. }
The first observation of SM CE$\nu$NS was achieved in 2017 by the COHERENT collaboration through an accelerator-based experiment~\cite{COHERENT:2017ipa}, confirmed by subsequent experiments~\cite{COHERENT:2020iec, COHERENT:2021xmm}, and the first indication of solar $^8$B neutrinos induced CE$\nu$NS events in the DM direct detection (DD) experiments was also reported by PandaX-4T~\cite{PandaX:2024muv} and XENONnT~\cite{XENON:2024ijk} in 2024.
\zbl{Low energy processes are less sensitive to the nuclear structure details~\cite{Vietze:2014vsa}, meaning that the nucleus can be approximated as a point particle characterized a charge and a spin. The nuclear effect can thus be encoded into spin-independent (SI) and spin-dependent (SD) form factors. In particular, the nuclear effect of the SM CE$\nu$NS process is well described by a SI form factor, characterizing the weak charge distribution within the nucleus~\cite{Freedman:1973yd}. 
The identical form factor was widely adopted in studies of neutrino-nucleus elastic scattering with tensor interactions~\cite{Barranco:2011wx, Lindner:2016wff, Papoulias:2017qdn, AristizabalSierra:2018eqm, AristizabalSierra:2019zmy, Li:2020lba, Chang:2020jwl, Han:2020pff, Demirci:2021zci, Flores:2021kzl, Majumdar:2021vdw, DeRomeri:2022twg, Chatterjee:2024vkd}. However, it has been known historically that tensor interactions contribute to SD interactions in low-energy processes, e.g., Gamow-Teller transitions in $\beta$ decay~\cite{Lee:1956qn}. Therefore, the nuclear effect of tensor interactions should be encapsulated into SD form factors, which was recognized very recently through a nuclear response framework~\cite{Altmannshofer:2018xyo, Hoferichter:2020osn}.}

\zbl{In the nuclear response framework, the nucleus is modeled by a collective nucleon system, leading to extra internal interactions between a point particle and a nucleus after the matching between relativistic interactions and non-relativistic (NR) nucleon operators~\cite{Fitzpatrick:2012ix, Anand:2013yka}. To describe the nuclear effect, this procedure adopts a complete set of six nuclear response functions instead of the conventional SI/SD form factors used in the elastic scattering. The nuclear response functions reveal the charge or current distribution inside the nucleus, depending on the underlying interactions and the corresponding nucleon operators. 
By matching the tensor interactions into the NR operators, previous studies~\cite{Altmannshofer:2018xyo, Hoferichter:2020osn} identified the leading-order operator in the momentum expansion as the spin operator $ \vec{\sigma}_N$. Consequently, the nuclear effect is encoded into two nuclear response functions of $ \vec{\sigma}_N$, resulting in a suppression of several orders of magnitude in the cross section compared to scenarios assuming a SI form factor. This significant suppression arises from the nuclear response functions, because the amplitude for a neutrino scattering off paired nucleons with opposite spin directions cancel with each other. This critical insight was overlooked for a long time but has recently been utilized in several phenomenological studies~\cite{Candela:2024ljb,DeRomeri:2024iaw,Chattaraj:2025rtj}.}

\zbl{However, we find that the spin-suppressed result also inadvertently overlooked other nucleon operators accompanied by a momentum suppression, which is quantified by the ratio of the momentum transfer to the nucleon mass $|\vec{q} |/m_N\sim\mathcal{O}(0.01)$. 
Specifically, concerning the unit operator $\mathrm{1}_N$ omitted in the spin-suppressed result, the amplitude for a neutrino scattering off a nucleus arises from the coherent sum of contributions from each individual nucleon. This coherent enhancement effect, scaling with the nuclear mass number, compensates the momentum suppression factor, and simultaneously amplifies the effect of the other operators through their interference with $\mathrm{1}_N$.}

In this work, we systematically investigate the tensor interactions in CE$\nu$NS in light of the nuclear response framework. The cross section can be decomposed into parity-odd and parity-even components, stemming from the nucleon operators suppressed by $|\vec{q} |/m_N$ and $\vec{\sigma}_N$, respectively. We find that the spin-suppressed result only contain the parity-even part, and the dominant contribution to the cross section originates from the parity-odd part, which exhibits no nuclear spin suppression and is coherently enhanced, thus yields a few orders of magnitude improvement over the spin-suppressed result.
We also discuss the sizable contribution to the CE$\nu$NS cross section from the neutrino magnetic moment ($\nu$MM) induced by a loop diagram of the tensor interactions~\cite{Papoulias:2015iga, Xu:2019dxe}.
In addition, we perform phenomenological investigations based on the data from the well-measured COHERENT experiment and recent solar $^8$B neutrino observations in the DM DD experiments at PandaX-4T and XENONnT.

{\bf Theoretical framework.} 
The neutral tensor interactions between quarks $\mathfrak{q}$ and neutrinos $\nu_\alpha$ can be described by:
\begin{equation}\label{eqn:Lagranian-quark}
\mathcal{L}_{\mathrm{eff}}^{T} =\sum _{\mathfrak{q}=u,d}\sum _{\alpha =e,\mu ,\tau } G_{T,\alpha }^{q } (\overline{\nu }_{\alpha } \sigma ^{\mu \nu } \nu _{\alpha } )(\overline{\mathfrak{q}} \sigma _{\mu \nu } \mathfrak{q})\,,
\end{equation}
with $\sigma^{\mu\nu}\equiv\frac i2[\gamma^\mu,\gamma^\nu]$. 
Nucleons, rather than quarks, serve as the effective hadronic degrees of freedom in describing low-energy scattering processes. Transitioning from quarks to nucleons can be achieved via the tensor interaction's nucleon matrix element \zbl{up to $\mathcal{O}(|\vec{q} |/m_N)$} \cite{Hoferichter:2020osn}:
\begin{small}
\begin{equation} \label{eqn:tensor-matrix-element}
\langle N_{p^{\prime }} |\overline{\mathfrak{q}} \sigma ^{\mu \nu } \mathfrak{q}|N_{p} \rangle =\overline{u}_{p'}\left[ \sigma ^{\mu \nu } F_{1,T}^{\mathfrak{q},N} -\frac{i}{m_{N}} (\gamma ^{\mu } q^{\nu } -\gamma ^{\nu } q^{\mu } )F_{2,T}^{\mathfrak{q},N}\right] u_{p}\,,
\end{equation}
\end{small}
where nucleon form factors evaluated at zero momentum transfer $F^{u,p}_{1,T}=0.784(28)$ and $F^{d,p}_{1,T}=-0.204(11)$~\cite{Gupta:2018lvp} are relative to the tensor charges, while $F^{u,p}_{2,T}=-1.5(1.0)$ and $F^{d,p}_{2,T}=0.5(3)$~\cite{Hoferichter:2018zwu} are relative to the tensor anomalous magnetic moments. Here, we assume the isospin symmetry: $F^{u,n}_{i,T}=F^{d,p}_{i,T}$ and $F^{d,n}_{i,T}=F^{u,p}_{i,T}~(i=1,2)$. Since the nucleon form factors can be approximated as constants within the energy range of CE$\nu$NS, the dependence on the momentum transfer is not explicitly included. 
\zbl{Here, kinematics of the neutrino-nucleon scattering is defined by: $\nu(k)+N(p)\to\nu(k^{\prime})+N(p^{\prime})$, with the momenta of the incoming (outgoing) neutrino and nucleon being $k$ ($k^\prime$) and $p$ ($p^\prime$), respectively, and the momentum transfer being $q\equiv p-p^{\prime}$.}
After suppressing the lepton flavor label for simplicity, we obtain the amplitude at the nucleon level \zbl{up to $\mathcal{O}(|\vec{q} |/m_N)$:
\begin{equation}\label{eqn:amplitude-nucleon}
\begin{aligned}
\mathcal{M}^{\mathrm{T}} =\sum _{N=p,n} & \overline{u}_{k'} \sigma _{\mu \nu } u_{k}\overline{u}_{p'}[{\displaystyle G_{T}^{N,1}} \sigma ^{\mu \nu }\\
 & -{\displaystyle G_{T}^{N,2}}\frac{i}{m_{N}} (\gamma ^{\mu } q^{\nu } -\gamma ^{\nu } q^{\mu } )] u_{p}
\end{aligned}\,,
\end{equation}
}with nucleon couplings $G_{T}^{N,i} \equiv \sum \limits_{\mathfrak{q}=u,d} G_{T}^{\mathfrak{q}} F_{i,T}^{\mathfrak{q},N} $ ($i=1,2$). 

In the nuclear response framework, the cross section is factorized into the leptonic and nuclear parts, where nuclear response functions are determined by non-relativistic (NR) nucleon operators. 
After performing the following NR expansions of nucleon-fermion bilinears, we can match Eqn.~\ref{eqn:amplitude-nucleon} with its NR form:
\begin{small}
\begin{equation}\label{eqn:nucleon-operator}
\mathcal{L}_{\mathrm{eff}}^{\mathrm{T,NR}} =\sum _{N=p,n} l_{0}^{N} 1_{N} +\vec{l}_{5}^{N} \cdot \vec{\sigma }_{N} +\vec{l}_{E}^{N} \cdot \left( -i\frac{\vec{P}_{N}}{2m_{N}} \times \vec{\sigma }_{N}\right)\,,
\end{equation}
\end{small}
where $\vec{P}_{N} $ is the total momentum operator, and $l_0^N$, $\vec{l}_5^N$ and $\vec{l}_E^N$ are the leptonic charges or currents associated with their corresponding operators. 
\zbl{According to the remaining spatial rotational symmetry~\cite{Glick-Magid:2022erc, Glick-Magid:2023uhk}, the tensor interactions are decomposed into three parts: $\mathcal{L}_{(0)}^{1}$, $\mathcal{L}_{(1)}^{1}$ and $\mathcal{L}_{(0)}^{2}$, where the superscripts denote their derivation from either the first or second term in Eqn.~\ref{eqn:amplitude-nucleon} and the subscripts label their underlying Lorentz structures. For instance, $G_{T}^{N,1}\overline{u}_{k'} \sigma ^{\mu \nu } u_{k}\overline{u}_{p'} \sigma _{\mu \nu } u_{p} =-2G_{T}^{N,1}\left(\overline{u}_{k'} \sigma _{(0)}^{i} u_{k}\overline{u}_{p'} \sigma _{(0)}^{i} u_{p} -\overline{u}_{k'} \sigma _{(1)}^{i} u_{k}\overline{u}_{p'} \sigma _{(1)}^{i} u_{p}\right) =\mathcal{L}_{(0)}^{1} +\mathcal{L}_{(1)}^{1}$, with $\sigma _{( 0)}^{i} =\sigma ^{0i}$, $\sigma _{( 1)}^{i} =\frac{1}{2} \epsilon ^{ijk} \sigma ^{jk}$.
The matching results are summarized in Tab.~\ref{tab:operator}, with the derivation details and explicit expressions of $l_0^N$, $\vec{l}_5^N$ and $\vec{l}_E^N$ provided in Appendix~\ref{supp:xsection}.
In this decomposition, the three terms are further decomposed into neutrino and nucleon components, and the parity of the nucleon operators is also determined based on their underlying Lorentz structures. 
The suppression of $1_N$ by $|\vec{q} |/m_N$ can be attributed to the necessity of preserving the parity. Specifically, the parity of the dimensionless quantities $\frac{\vec{P}_N}{m_N}$, $\vec{\sigma}_N$ and $\frac{i\vec{q}}{m_N}$ are odd, even and odd, respectively, aligning with their relativistic Lorentz structures.
Additionally, the suppression can also be geometrically interpreted as the rapidity of boosting the initial nucleon to a scattered nucleon.
}
\zbl{Note that our analysis in this work is confined to one-body operators. Two-body currents can be converted into one-body operators associated with indeterminate low energy constants, warranting further exploration in future studies; see Appendix~\ref{supp:2b} for more details.}

\begin{table}[t]\label{tab:operator}
\caption{NR expansions of the tensor interactions up to $\mathcal{O}(|\vec{q} |/m_N)$, where the nuclear state normalization factor $\rho \simeq 2M$ with $M$ being the nucleus mass.}
\centering
\begin{tabular}{|c|c|c|c|}
\hline 
    & $\mathcal{L}_{(0)}^{1}$ &  $\mathcal{L}_{(1)}^{1}$ & $\mathcal{L}_{(0)}^{2}$ \\
\hline 
 neutrino & $-2\overline{u}_{k ^{\prime }}\vec{\sigma }_{(0)} u_{k }$ & $2\overline{u}_{k ^{\prime }}\vec{\sigma }_{(1)} u_{k }$ & $-2\overline{u}_{k ^{\prime }}\vec{\sigma }_{(0)} u_{k }$ \\
\hline 
 nucleon &  $\frac{i\vec{q}1_{N} -\vec{P} \times \vec{\sigma}_{N}}{2m_{N}} G_{T}^{N,1}\rho$ & $\vec{\sigma}_{N} G_{T}^{N,1}\rho$ & $-\frac{i\vec{q}}{m_{N}} 1_{N} G_{T}^{N,2}\rho$ \\
 \hline
 parity& odd& even&odd\\\hline
\end{tabular}
\end{table}

Since the nucleus is modeled as a composite system, the nucleon operator $O_N$ is accompanied by an additional spatial operator $e^{-i\vec{q}\cdot\vec{x}}$~\cite{Anand:2013yka}. 
The factor $e^{-i\vec{q}\cdot\vec{x}}$ can be done by performing a multipole expansion~\cite{Donnelly:1979ezn,Haxton:2008zza}, obtained by expanding the exponential into a series of terms that give rise to spherical tensor operators. The transition reduced nuclear matrix element for the multipole operator, i.e., the nuclear response function, can be evaluated by the large scale shell model.
For example, the multipole operator of $1_{N}$ is denoted by $M_J$, where $J$ represents its rank, leading to the nuclear response function $\mathcal{F}^{M_J}_N(|\vec{q} |^2)$. 
At the long-wavelength limit, $1_{N}$ counts the number of nucleons inside the nucleus, leading to $\mathcal{F}_p^{M_0}(0)=Z$ and $\mathcal{F}_n^{M_0}(0)=N$ with the proton number $Z$ and the neutron number $N$. 
With the aid of the parity and time reversal properties of multipole operators, we find that only four nuclear response functions remain in our discussion~\cite{Fitzpatrick:2012ix,Anand:2013yka, DelNobile:2021wmp}: $\mathcal{F}^{M}_N$ from $1_{N}$, $\mathcal{F}^{\Sigma^\prime}_N$ and $\mathcal{F}^{\Sigma^{\prime \prime}}_N$ from the spin operator $\vec{\sigma }_{N}$, and $\mathcal{F}_{N}^{\Phi ^{\prime \prime }}$ from the spin-orbit operator $\frac{\vec{P}_N \times \vec{\sigma}_{N}}{2m_{N}}$, where the nuclear response function is now denoted by $\mathcal{F}^{O}_N$ for simplicity.
For an evaluation of nuclear response functions, please refer to Appendix~\ref{supp:response}.

{\bf Cross sections.} 
The cross section of the tensor interactions can be derived as in Ref.~\cite{Anand:2013yka}, with further details given in Appendix~\ref{supp:xsection}. We find that the differential cross section in terms of the recoil energy $E_r$ is decomposed into P-odd (O) and P-even (E) terms: 
\begin{equation}\label{eqn:cross_section}
\frac{d\sigma _{T}}{dE_r} =\frac{d\sigma _{T}^{\mathrm{O}}}{dE_r} +\frac{d\sigma _{T}^{\mathrm{E}}}{dE_r}\,,
\end{equation}
where 
\begin{equation}\label{eqn:cross-section-odd}
\begin{aligned}
\frac{d\sigma _{T}^{\mathrm{O}}}{dE_r} = & \frac{M}{2\pi }\frac{|\vec{q}|^{2}}{m_{N}^{2}}\sum _{N,N^{\prime } =p,n}\left( c_{0}^{N} c_{0}^{N^{\prime }}\mathcal{F}_{N}^{M}\mathcal{F}_{N^{\prime }}^{M}\right. \\
+4{\displaystyle c_{0}^{N} G_{T}^{N^{\prime } ,1}} & \left.  \mathcal{F}_{N}^{\Phi ^{\prime \prime }}\mathcal{F}_{N^{\prime }}^{M} +4{\displaystyle G_{T}^{N,1} G_{T}^{N^{\prime } ,1}}\mathcal{F}_{N}^{\Phi ^{\prime \prime }}\mathcal{F}_{N^{\prime }}^{\Phi ^{\prime \prime }}\right)
\left( 1-\frac{E_r}{E_{\nu }}\right)
\end{aligned}\,,
\end{equation}
with $c_{0}^{N} \equiv -G_{T}^{N,1} +2G_{T}^{N,2}$ and
\begin{equation}\label{eqn:cross-section-even}
\begin{aligned}
\frac{d\sigma _{T}^{\mathrm{E}}}{dE_r} = & \frac{M}{\pi }\sum _{N,N^{\prime } =p,n}[2{\displaystyle G_{T}^{N,1} G_{T}^{N^{\prime } ,1} \left( 1-\frac{E_r}{E_{\nu }}\right)\mathcal{F}_{N}^{\Sigma ^{\prime \prime }}\mathcal{F}_{N^{\prime }}^{\Sigma ^{\prime \prime }}}\\
 & +{\displaystyle G_{T}^{N,1} G_{T}^{N^{\prime } ,1}}\left( 1-\frac{ME_r}{2E_{\nu }^{2}} -\frac{E_r}{E_{\nu }}\right)\mathcal{F}_{N}^{\Sigma ^{\prime }}\mathcal{F}_{N^{\prime }}^{\Sigma ^{\prime }}]
\end{aligned}\,.
\end{equation}
The P-odd cross section is dominated by the $\mathcal{F}_{N}^{M}\mathcal{F}_{N^{\prime }}^{M}$ term, while the $\mathcal{F}_{N}^{\Phi ^{\prime \prime }}\mathcal{F}_{N^{\prime }}^{M}$ term, arising from the interference between $1_N$ and the spin-orbit operator, constitutes a sizable secondary contribution.
This is in contrast to the vector case, where the contribution of the spin-orbit operator can be neglected, e.g., in the evaluation of the weak radius~\cite{Hoferichter:2020osn,Coloma:2020nhf}. 
\zbl{Notice that $F^{\mathfrak{q},N}_{2,T}$ and $F^{\mathfrak{q},N}_{1,T}$ are of the same order in $c_{0}^{N}$, leading to comparable contributions from the tensor charge and the tensor anomalous magnetic moment.}
Qualitatively, the nuclear response functions for P-odd and P-even terms are estimated by the mass number $A$ and the spin expectation value $\langle\mathbf{S}_{N}\rangle$ based on their normalizations.
The ratio between the two parts can be estimated by $\sigma _{T}^{\mathrm{O}}/\sigma _{T}^{\mathrm{E}}\simeq\frac{A^2|\vec{q} |^2}{m_N^2 \langle\mathbf{S}_{N}\rangle^2}\gtrsim \mathcal{O}(10)$ for a heavy nucleus with $A\gtrsim 100$, given $|\vec{q} |\sim 10~\mathrm{MeV}$, $m_N\sim 1~\mathrm{GeV}$ and $\langle\mathbf{S}_{N}\rangle^2\sim \mathcal{O}(0.1)$~\cite{Fitzpatrick:2012ix, Klos:2013rwa,Hu:2021awl}.
\zbl{Consequently, the P-odd part tends to dominate over the P-even part due to its spin-dependent suppression. 
We also verify that the results on the tensor interactions in Refs.~\cite{Altmannshofer:2018xyo,Hoferichter:2020osn} align with our P-even cross section; see Appendix~\ref{supp:check} for further details.}

\begin{figure}
    \centering
    \includegraphics[width=0.9\linewidth]{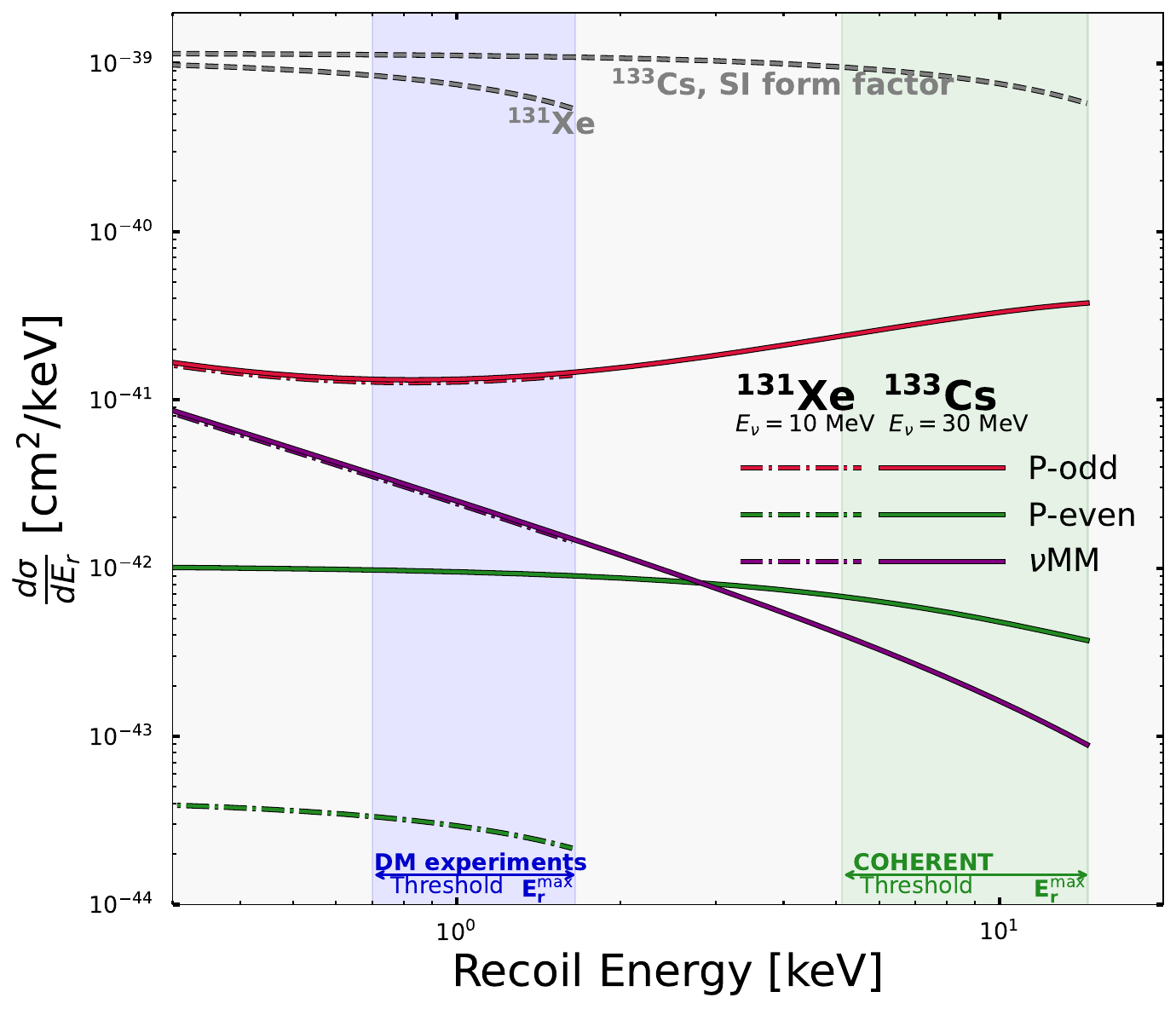}
    \caption{\zbl{The differential cross section for $^{133}$Cs and $^{131}$Xe with incident neutrino energies at 30~$\mathrm{MeV}$ and 10~$\mathrm{MeV}$, assuming $G_{T}^u=G_{F}$ and $G_{T}^d=0$. The incorrect results (gray dashed lines) calculated using SI form factors are presented for comparative purposes~\cite{AristizabalSierra:2018eqm}.}}
    \label{fig:cross-section}
\end{figure}

In addition to the tree-level contributions discussed above, the tensor interactions can naturally induce a neutrino magnetic moment ($\nu$MM)~\cite{Papoulias:2015iga, Xu:2019dxe}:
\begin{equation}
\mu _{\nu }^{\mathfrak{q}} \approx \frac{m_{e} N_{c}}{\pi ^{2}} G_{T}^{\mathfrak{q}} m_{\mathfrak{q}} Q_{\mathfrak{q}} [1+\ln (G_{T}^{\mathfrak{q}} m_{\mathfrak{q}}^{2}) ]\mu _{B}\,,
\end{equation}
where $N_c$ is the number of quark colors, $Q_\mathfrak{q}$ is the quark electric charge, $\mu_B$ is the Born magnetic moment, $m_e$ and $m_\mathfrak{q}$ are the electron and quark masses, respectively. The sign of $\nu$MM is flipped for the antineutrino. Note that the tensor interaction induced $\nu$MM (TI-$\nu$MM) depends on the UV behavior of the effective interaction, and could be canceled by other mechanisms that generate $\nu$MM~\cite{Giunti:2014ixa, Giunti:2024gec}. It is worth highlighting that TI-$\nu$MM would vanish if the neutrino had similar tensor couplings with $u$ and $d$ quarks, due to the cancellation resulting from the opposite electric charges of the two quarks. 
If we take account of the TI-$\nu$MM, $\frac{d\sigma _{T}^{\mathrm{O}}}{dE_r}$ in Eqn.~\ref{eqn:cross-section-odd} is modified with the following replacement:
\begin{equation} \label{eqn:c0}
c_{0}^{p}\rightarrow -G_{T}^{p,1} +2G_{T}^{p,2} +\frac{m_{N}}{|\vec{q} |^{2}} e\sum _{\mathfrak{q}=u,d} \mu _{\nu }^{\mathfrak{q}}\,,
\end{equation}
where $e$ is the electromagnetic coupling constant. Thus, the quadratic contribution of TI-$\nu$MM is proportional to $1/|\vec{q} |^2 \sim  1/E_r$, which dominates at low energies. 

Taking $^{133}$Cs and $^{131}$Xe as benchmark targets, the differential cross section for the tensor interactions with $G_{T}^u=G_{F}$ is illustrated in Fig.~\ref{fig:cross-section} with $G_{F}$ being the Fermi constant. The benchmark neutrino energies are $30~\mathrm{MeV}$ and $10~\mathrm{MeV}$, for a typical muon neutrino energy in the COHERENT experiment and $^8$B neutrino energy in the DM experiment, respectively. It is evident that the P-odd contribution predominates the cross section, which is approximately two and three orders of magnitude larger than that from the spin-suppressed result for $^{133}$Cs and $^{131}$Xe, respectively.
\zbl{The contribution of $\nu$MM is also presented. Since it interferes with the tree-level contributions, as one can see from Eqn.~\ref{eqn:c0}, we eliminate the tree-level contributions by setting $G^{N,i}_T=0$ in Eqn.~\ref{eqn:cross_section} to show the contribution of $\nu$MM alone.} The absence of the suppression $|\vec{q} |/m_N$ makes $\nu$MM pronounced for low energy threshold experiments such as DM direct detection experiments. The opposite tensor quark couplings, e.g., $G_{T}^u=-G_{T}^d=G_{F}$, result in a large $\nu$MM and its overwhelming effect.

\begin{figure}
    \centering
    \includegraphics[width=0.9\linewidth]{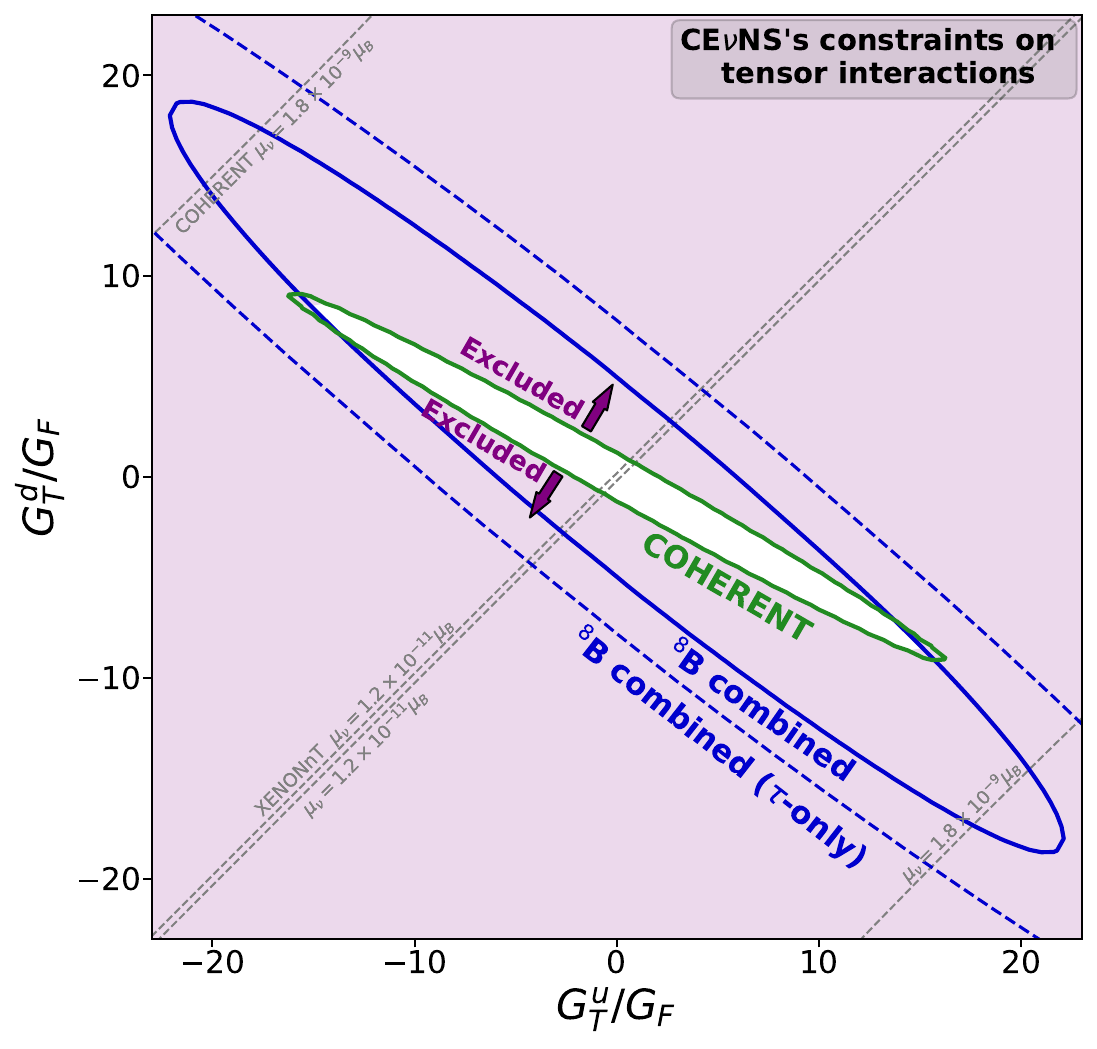}
    \caption{CE$\nu$NS's constraints (light purple regions) on neutrino flavor-independent tensor couplings: $G^u_T$ and $G^d_T$. Blue dashed line represents the constraints on $G^u_{T,\tau}$ and $G^d_{T,\tau}$ for only $\tau$-flavor couplings being nonzero, and the axis labels should be modified accordingly. $^8$B combined represents the constraints from the combined data analysis of PandaX-4T and XENONnT.}
    \label{fig:constraints}
\end{figure}

{\bf Experimental constraints.} Based on the COHERENT-2021 data~\cite{COHERENT:2021xmm} and $^8$B neutrino observations in the DM direct detection experiments~\cite{PandaX:2024muv,XENON:2024ijk}, we present the 90\% confidence level (CL) constraints on the tensor interactions, as illustrated in Fig. \ref{fig:constraints}. For further details on the spectra and the statistical method, please refer to Appendix~\ref{supp:analysis}.
The COHERENT-2021 data~\cite{COHERENT:2021xmm} set the strongest limits on lepton-flavor-independent tensor couplings for CE$\nu$NS experiments. Due to the interference of proton and neutron couplings in Eqn.~\ref{eqn:cross_section}, the cross section diminishes along the direction of the diagonal strip in Fig. \ref{fig:constraints}, leading to the formation of the survival strip. \zbl{Moreover, the constraints are anticipated to be improved by approximately one order of magnitude in future COHERENT experiments~\cite{COHERENT:2022nrm}.}
In comparison, the large statistical uncertainties in the DM experiments lead to a less stringent constraint. However, the neutrino oscillation offers a valuable channel to probe $\tau$-flavor tensor couplings, which are inaccessible through the COHERENT experiment. In this context, the neutrino propagation from the Sun to Earth and the matter effect must be included. The relevant result is also presented in Fig. \ref{fig:constraints}, which might facilitate studies on neutrino-related new physics within the neutrino fog~\cite{OHare:2021utq,Tang:2023xub, Tang:2024prl,Blanco-Mas:2024ale}. 

Moreover, TI-$\nu$MM can be probed by alternative experiments~\cite{Giunti:2014ixa, Giunti:2024gec}. Due to its dependence on the energy scale of new physics, a comprehensive examination of $\nu$MM is beyond the scope of this work.
For illustration, we present two conservative limits on $\nu$MM derived from COHERENT~\cite{Coloma:2022avw,AtzoriCorona:2022qrf} and XENONnT~\cite{Khan:2022bel,A:2022acy}  experiments. However, due to the opposite electric charges of $u$ and $d$ quarks, TI-$\nu$MM is expected to vanish in the universal coupling scenario. The skew diagonal strip enclosed by two dashed lines in Fig. \ref{fig:constraints}  is permitted by the limits of $\nu$MM but excluded by CE$\nu$NS experiments, thereby indicating that CE$\nu$NS experiments complement $\nu$MM in probing the tensor interactions.

Furthermore, the stringent constraint on tensor interactions arises from the upper limit on the neutrino mass~\cite{Prezeau:2004md,Ito:2004sh}, which also depends on the energy scale of new physics. This constraint might be alleviated by incorporating pseudo-scalar interactions or alternative neutrino mass generation mechanisms. Similarly, CE$\nu$NS experiments can exclude the parameter space with opposite couplings for $u$ and $d$ quarks, where the tensor interaction-induced neutrino masses are too small to be constrained by the upper limit on the neutrino mass.

\begin{figure}
    \centering
    \includegraphics[width=0.9\linewidth]{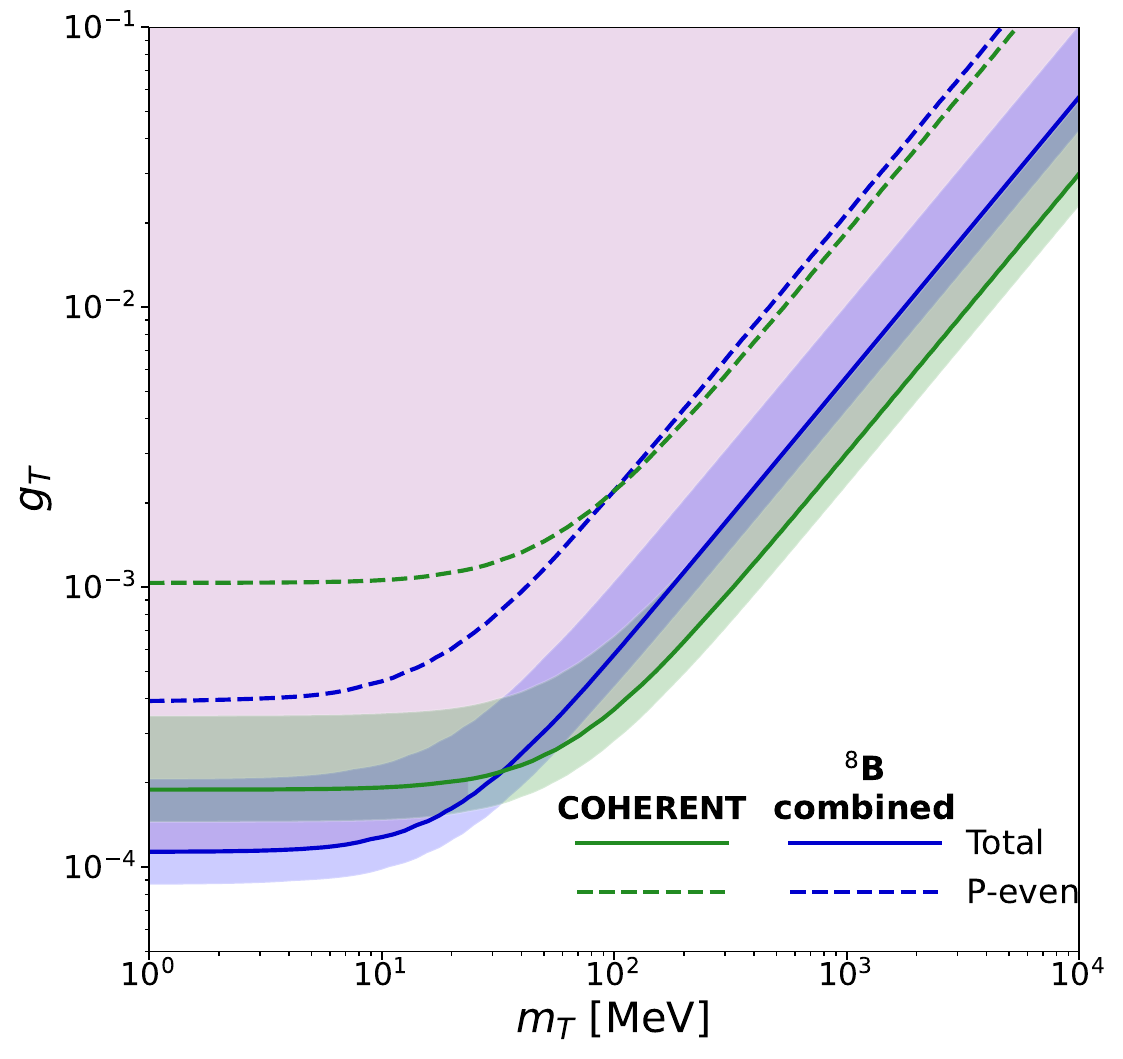}
    \caption{Constraints on the universal tensor interaction with a light mediator from CE$\nu$NS experiments.}
    \label{fig:constraints-light}
\end{figure}

The universal tensor interaction with a light mediator is also demonstrated. To generalize to the light mediator scenario, one can substitute the following replacement in Eqn.~\ref{eqn:Lagranian-quark}:
\begin{equation}\label{eqn:propagator}
G_{T,\alpha }^{\mathfrak{q}}\rightarrow \frac{g_{T}^{2}}{m_{T}^{2} +|\vec{q} |^{2}}\,,
\end{equation}
where $g_T$ and $m_T$ are the universal tensor coupling and the light mediator mass, respectively. As illustrated in Fig.~\ref{fig:constraints-light}, the green and blue lines delimit the boundaries of the 90\% CL excluded regions derived from the COHERENT-2021 data and the DM experiments, respectively. In the high-mass region, the propagator is uniquely determined by $m_T$ since $|\vec{q}|\ll m_T$, leading to diagonal lines proportional to $m_T$. Conversely, in the low-mass region, the flattens of the boundaries come from the fact that $|\vec{q}|\gg m_T$, indicating that the scattering experiments lose their sensitivities to lighter mediators. DM detectors, with a lower energy threshold, are more sensitive to the smaller momentum transfer compared to COHERENT experiments. Therefore, the DM experiments put more stringent constraints on $m_T\lesssim 30~\mathrm{MeV}$. It is undisputed that two types of CE$\nu$NS experiments are complementary in probing neutrino-related new physics. 
Moreover, we demonstrate the effect of the large uncertainty on nucleon form factors. For example, we vary $F^{u,p}_{2,T}$ within its 1$\sigma$ region, presenting the results as shaded bands in Fig.~\ref{fig:constraints-light}. The uncertainties of nucleon form factors, particularly for $F^{u,p}_{2,T}$ and $F^{d,p}_{2,T}$, substantially affect the CE$\nu$NS constraints on the tensor interactions. These effects could be mitigated by a more precise evaluation of nucleon form factors in the future. For comparison, we also present the results for the spin-suppressed result, which are considerably less stringent than those of the full contribution.

{\bf Summary.} 
In this work, we pointed out that the conventional SI form factor is insufficient to describe the nuclear effect for the tensor interactions. Utilizing the nuclear response framework, \zbl{we analyzed the leading contribution for the cross section of the tensor interactions that have been overlooked in previous studies.} We found that the anomalous magnetic moment of the tensor plays a role comparable to the tensor charge. Our result revealed that the tensor interactions evade the nuclear spin suppression, leading to an enhancement to the cross section by a few orders of magnitude. We systematically incorporated previously omitted TI-$\nu$MM into our discussion.

Employing COHERENT data and recent solar neutrino observations in DM detectors, we demonstrated the complementarity of constraints from CE$\nu$NS and $\nu$MM on the tensor interactions. We also showed that novel solar neutrino observations advance in probing tensor interactions with $\tau$-flavor couplings and a light mediator. Future higher-precision CE$\nu$NS experiments will undoubtedly improve sensitivities in the search of new physics. For tensor interactions, it is also crucial to improve the evaluation of the nucleon form factor. 
Moreover, our results motivated revised interpretations on CE$\nu$NS experiments using spin-zero nuclei.

{\it Acknowledgments.} 
We appreciate Gang Li, Ning-Qiang Song, Feng-Jie Tang and Jiang-Hao Yu for fruitful discussions. J.T. is supported in part by National Natural Science Foundation of China under Grant Nos. 12347105 and 12075326. J.L. is supported by the National Natural Science Foundation of China under Grant Nos.~12275368 and the Fundamental Research Funds for the Central Universities, Sun Yat-Sen University under Grant No. 24qnpy116. This work was supported in part by Fundamental Research Funds for the Central Universities (23xkjc017) in Sun Yat-sen University. J.T. is grateful to Southern Center for Nuclear-Science Theory (SCNT) at Institute of Modern Physics in Chinese Academy of Sciences for hospitality.

\twocolumngrid
\section*{References}
\def\bibsection{}
\bibliography{reference}

\clearpage
\onecolumngrid
\appendix 

\section{Nuclear response}
\label{supp:response}
To evaluate the nuclear response functions, one needs to deal with the many-body calculation. Fortunately, the traditional non-relativistic many-body description of the nucleus has been developed in Ref.~\cite{Fetter1971QuantumTO}. 
The nucleus is modeled as a quantum mechanical system of point nucleons, where each nucleon is described by a single-particle Hartree-Fock state with a definite nodal number and angular momentum. For a single-nucleon multipole operator $O_{N}^{J}$, which is an irreducible tensor operator with rank $J$, the transition reduced nuclear matrix element between initial and final nuclear many-body states $\Psi _{i}$ and $\Psi _{f}$ reads~\cite{OConnell:1972edu,Walecka1995TheoreticalNA}:
\begin{equation}
\langle \Psi _{f} ||\hat{O}_{N}^{J} ||\Psi _{i} \rangle =\sum _{\alpha ,\beta } \langle \alpha ||O_{N}^{J} ||\beta \rangle \psi _{\alpha \beta }^{fi}\,,
\end{equation}
where $\hat{O}_{N}^{J}$ is the second quantization version of $O_{N}^{J}$, and $|\alpha \rangle, $ $|\beta \rangle $ represents the single-particle Hartree-Fock state. The single-nucleon matrix element $\langle \alpha ||O_{N}^{J} ||\beta \rangle $ can be evaluated analytically in a harmonic oscillator single-particle basis, yielding a form like $e^{-\frac{u}{2}}p(u)$, where $p(u)$ is a polynomial in $u = |\vec{q}|^2b^2/2$ with the oscillator parameter $b$ and the three-momentum transfer $|\vec{q}|$~\cite{Donnelly:1979ezn,Donnelly:1984rg,Haxton:2008zza}.   The one-body density matrix element $\psi _{\alpha \beta }^{fi}$, generated by performing a large-scale nuclear shell model calculation, encodes the many-body physics. For isotopes ($^{127}$I, $^{133}$Cs, $^{131}$Xe, etc.)  discussed in this work, we adopt GCN5082~\cite{Menendez:2008jp} as the shell model interaction.
The calculation of the reduced nuclear matrix element can be readily generated by $\mathbf{dmformfactor}$~\cite{Anand:2013yka}~\footnote{ Updated single nucleon density matrix elements can be found in $\mathbf{dmscatter}$\cite{Gorton:2022eed} and \url{https://github.com/Berkeley-Electroweak-Physics/Elastic}. }.
In CE$\nu$NS, we define the nuclear response  function: 
\begin{equation}
\mathcal{F}_{N}^{O^{J}_N} (\mathbf{q}^{2} )\equiv \sqrt{\frac{4\pi }{2J_{i} +1}} \langle ||\hat{O}_{N}^{J} ||\rangle =\sqrt{\frac{4\pi }{2J_{i} +1}} e^{-\frac{u}{2}}\sum _{j} c_{j}^{O^{J}_N} u^{j}\,,
\end{equation}
where $|\rangle$ is the nuclear ground state, $J_i$ is the angular momentum of the ground state and the coefficients $c_{j}^{O^{J}_N}$ can be derived analytically. We always suppress the rank label and the nuclear response function is denoted by $\mathcal{F}^{O}_N(\mathbf{q}^2)$ for simplicity. 
In addition, we provide a python script for the evaluation of the nuclear response functions and the cross section of the tensor interactions~\footnote{\url{https://github.com/zhangblong/CEvNSTensor}.}. 

\section{Details of cross section calculations}
\label{supp:xsection}
\begin{figure}[h]
    \centering
    \includegraphics[width=0.5\linewidth]{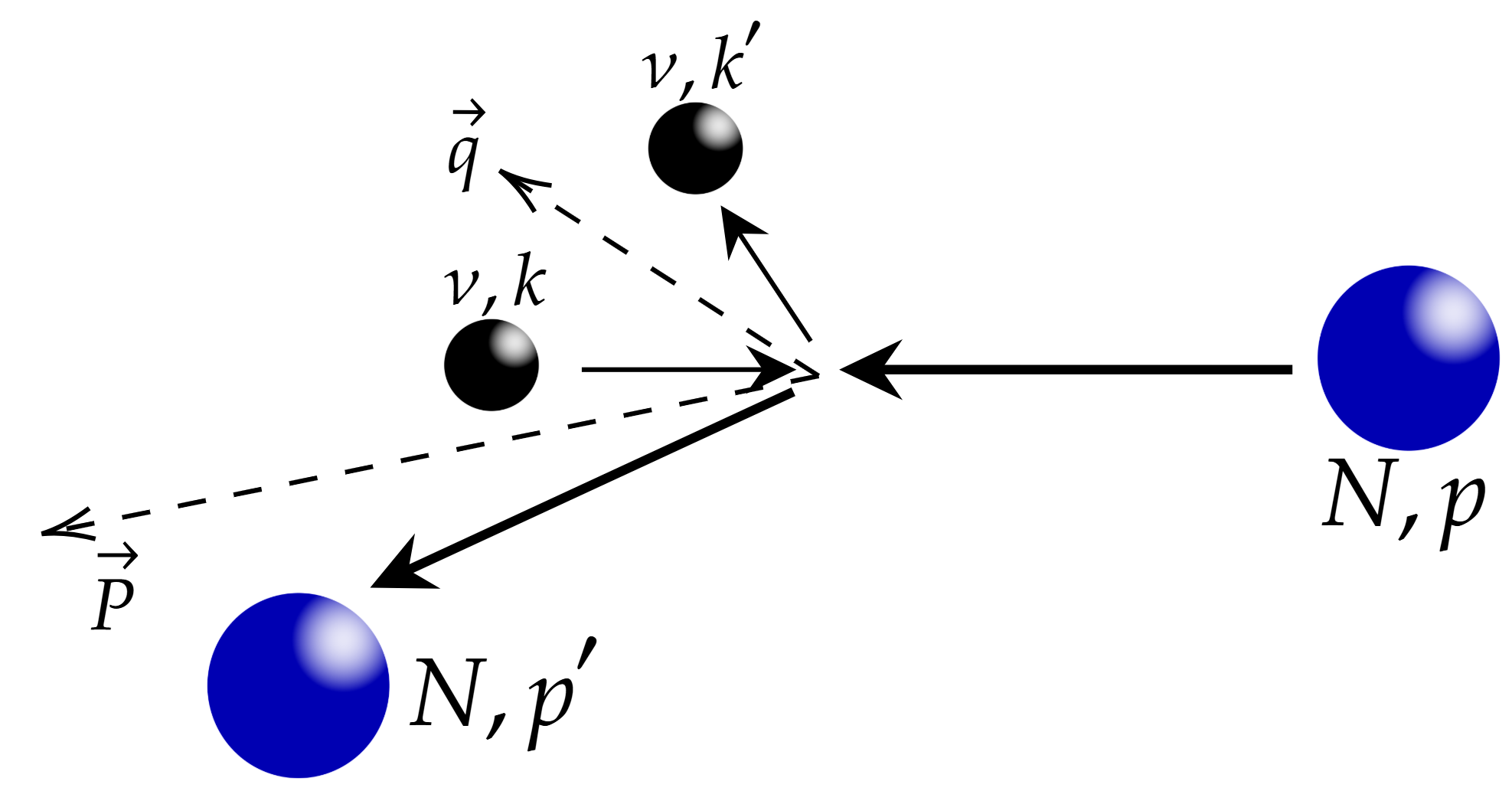}
    \caption{\zbl{A sketch of the kinematics in the neutrino-nucleon elastic scattering process.}}
    \label{fig:sketch}
\end{figure}
The procedure of the NR expansion on the tensor interactions is now introduced.
\zbl{As illustrated in Fig.~\ref{fig:sketch}, kinematics of the neutrino-nucleon elastic scattering is defined by: $\nu(k)+N(p)\to\nu(k^{\prime})+N(p^{\prime})$, with the momentum transfer $q\equiv k^{\prime}-k=p-p^{\prime}$.}
At the NR limit, the Weyl spinor for a nucleon reads: $u^{s} (p)=\frac{1}{\sqrt{4m_N}}\begin{pmatrix}
(2m_N-\vec{p} \cdot \vec{\sigma } )\xi ^{s}\\
(2m_N+\vec{p} \cdot \vec{\sigma } )\xi ^{s}
\end{pmatrix}$, with a helicity $s$~\footnote{We following the convention in Ref.~\cite{DelNobile:2021wmp}, while some literature~\cite{Anand:2013yka, Hoferichter:2020osn} adopt Bjorken and Drell $\gamma$ matrix conventions and spinor normalization (1 instead of the 2$m_N$).}. Thus, nucleon fermion bilinears $\bar{u}(p^{\prime})\Gamma u(p)$ can be computed accordingly, where $\Gamma$ represents the Lorentz structure of the interaction and the helicity $s$ is suppressed. 
According to the spatial rotational symmetry, we have $\overline{u}_{k'} \sigma ^{\mu \nu } u_{k}\overline{u}_{p'} \sigma _{\mu \nu } u_{p} =-2\left(\overline{u}_{k'} \sigma _{(0)}^{i} u_{k}\overline{u}_{p'} \sigma _{(0)}^{i} u_{p} -\overline{u}_{k'} \sigma _{(1)}^{i} u_{k}\overline{u}_{p'} \sigma _{(1)}^{i} u_{p}\right)$, with $\sigma _{( 0)}^{i} =\sigma ^{0i} =-i\begin{pmatrix}
\sigma _{i} & 0\\
0 & -\sigma _{i}
\end{pmatrix} $ and $\sigma _{( 1)}^{i} =\frac{1}{2} \epsilon ^{ijk} \sigma ^{jk}=\begin{pmatrix}
\sigma _{i} & 0\\
0 & \sigma _{i}
\end{pmatrix}$. 
For the second term in Eqn.~\ref{eqn:amplitude-nucleon}, we have: $\overline{u}_{k'} \sigma ^{\mu \nu } u_{k}\overline{u}_{p'} (\gamma ^{\mu } q^{\nu } -\gamma ^{\nu } q^{\mu } )u_{p} \simeq -2\overline{u}_{k'} \sigma _{(0)}^{i} u_{k}\overline{u}_{p'} q^{i} \gamma ^{0} u_{p}$. Other terms are suppressed by $\mathcal{O}((|\vec{q} |/m_N)^2)$, which can be safely neglected. 
The results of the NR expansion are listed in Tab.~\ref{tab:operator}.
Then one can read out the leptonic components in Eqn.~\ref{eqn:nucleon-operator}:
\begin{equation}\label{eqn:currents}
  \begin{array}{l}
l_{0}^{N} =\overline{u}_{k^{\prime }}\Biggl[ -2\Bigl( G_{T}^{N,1} -2G_{T}^{N,2}\Bigr)\frac{i|\vec{q}|}{2m_{N}}\Biggr] \sigma _{(0)}^{3} u_{k} \rho ,\\
\vec{l}_{5}^{N} =\overline{u}_{k^{\prime }}\Bigl( 2G_{T}^{N,1}\vec{\sigma }_{(1)}\Bigr) u_{k} \rho ,~\vec{l}_{E}^{N} =\overline{u}_{k^{\prime }}\Bigl( 2iG_{T}^{N,1}\vec{\sigma }_{(0)}\Bigr) u_{k} \rho\,,
\end{array}
\end{equation}
\zbl{where $\rho$ is the nuclear state normalization factor, arising from the fact that an additional factor $\frac{2M}{2m_N}$ is necessitated when we consider the transformation between quantized volumes of a nuclear state and a nucleon state~\cite{DelNobile:2021wmp}.}
\zbl{In addition, the suppression $|\vec{q} |/m_N$ in $l^N_0$ can be geometrically interpreted as the rapidity of boosting $p$ to $p^\prime$:
\begin{equation}
\eta \simeq \sinh \eta =\frac{|\vec{p} |\sqrt{m_{N}^{2} +|\vec{p} -\vec{q} |^{2}} -|\vec{p} -\vec{q} |\sqrt{m_{N}^{2} +|\vec{p} |^{2}}}{m_{N}^{2}} \simeq \frac{|\vec{q} |}{m_{N}}\,,
\end{equation}
where we assume $\vec{q}$ aligns with $\vec{p}$ for simplicity and $|\vec{q} |\ll |\vec{p} |\ll m_{N}$.}

Through the matching between the nucleon operators and the nuclear response functions, the averaged squared amplitude $\overline{|\mathcal{M} |^{2}}$ for Eqn.~\ref{eqn:nucleon-operator} reads~\cite{Anand:2013yka}:
\begin{equation}\label{eqn:amplitude}
\begin{aligned}
\overline{|\mathcal{M}|^{2}} = & \sum _{N,N^{\prime }}\{l_{0}^{N} l{_{0}^{N^{\prime }}}^{*}\mathcal{F}_{N}^{M}\mathcal{F}_{N^{\prime }}^{M} +\frac{|\vec{q}|^{2}}{m_{N}^{2}} l_{E,3}^{N} l{_{E,3}^{N^{\prime }}}^{*}\mathcal{F}_{N}^{\Phi ^{\prime \prime }}\mathcal{F}_{N^{\prime }}^{\Phi ^{\prime \prime }} +2\frac{|\vec{q}|}{m_{N}} Re\left[ l_{0}^{N} l{_{E,3}^{N^{\prime }}}^{*}\right]\mathcal{F}_{N}^{\Phi ^{\prime \prime }}\mathcal{F}_{N^{\prime }}^{M}\\
 & l_{5,3}^{N} l_{5,3}^{N^{\prime } *}{\displaystyle \mathcal{F}_{N}^{\Sigma ^{\prime \prime }}\mathcal{F}_{N^{\prime }}^{\Sigma ^{\prime \prime }}} +\frac{1}{2}\left(\vec{l}_{5}^{N} \cdot \vec{l}_{5}^{N^{\prime } *} -l_{5,3}^{N} l_{5,3}^{N^{\prime } *}\right)\mathcal{F}_{N}^{\Sigma ^{\prime }}\mathcal{F}_{N^{\prime }}^{\Sigma ^{\prime }}\}\,,
\end{aligned}
\end{equation}
where the transverse contribution of $\vec{l}_{E}^{N}$, proportional to $\mathcal{F}_{N}^{\tilde{\Phi }^{\prime }}\mathcal{F}_{N^{\prime }}^{\tilde{\Phi }^{\prime }}$, is neglected.
Recovering the rank label, terms like $\mathcal{F}_{N}^{M}\mathcal{F}_{N^{\prime }}^{M}$ should be understood as the incoherent sum of the contributions of the multipole operators with different $J$: $\mathcal{F}_{N}^{M}\mathcal{F}_{N^{\prime }}^{M}=\sum_J\mathcal{F}_{N}^{M_J}\mathcal{F}_{N^{\prime }}^{M_J}$, since operators with different $J$ are orthogonal. The rank of multipole operators must satisfy a triangular in-equation $0\leq J \leq 2J_i$ from the Wigner-Eckart theorem. In the low-energy process, rank $0$ operators without a $\mathcal{O}((|\vec{q} |/m_N)^J)$ suppression always dominate.
The next is to evaluate the leptonic parts in Eqn.~\ref{eqn:amplitude}. \zbl{For example, considering the incident left-handed neutrino, we have
\begin{equation}\label{eqn:l0derivation}
l_{0}^{N} l{_{0}^{N^{\prime }}}^{*} =c_{0}^{N} c_{0}^{N^{\prime }}\frac{|\vec{q} |^{2}}{m_{N}^{2}}  \mathrm{Tr}\left[\not{k}^\prime \sigma ^{03} P_{L}\not{k} \sigma ^{03}\right] =c_{0}^{N} c_{0}^{N^{\prime }}\frac{|\vec{q} |^{2}}{m_{N}^{2}} 2k^{0} k^{\prime 0}( 1+\hat{k} \cdot \hat{k}^\prime -2\hat{k} \cdot \hat{q}\hat{k}^\prime \cdot \hat{q})\,,
\end{equation}
where $\hat{k}_{i} \equiv \vec{k}_{i} /k_{i}^{0}$, and $\hat{q} \equiv \vec{q} /|\vec{q} |=\left( k^{\prime 0}\hat{k}^\prime -k^{0}\hat{k}\right) /|\vec{q} |$. Assuming the initial nucleus at rest, one can obtain $|\vec{q} |^{2} =2ME_{r}$ and $E_{r} =\frac{ME_{\nu }^{2}}{M^{2} +2ME_{\nu }}( 1-\cos \theta )$, where $\cos \theta \equiv \hat{k} \cdot \hat{k}^\prime$. Through a tedious calculation, the last term in Eqn.~\ref{eqn:l0derivation} can be solved by:
\begin{equation}
\hat{k} \cdot \hat{q}\hat{k}^\prime \cdot \hat{q} =\frac{\left( k^{\prime 0}\cos \theta -k^{0}\right)\left( k^{\prime 0} -k^{0}\cos \theta \right)}{|\vec{q} |^{2}} =\frac{1}{2}(\cos \theta -1)\,,
\end{equation}
where the relation $\left( q^{0}\right)^{2} =\left( k^{0}\right)^{2} -2k^{0} k^{\prime 0} +\left( k^{\prime 0}\right)^{2} \approx 0$ has been used. Consequently, up to $\mathcal{O}(E_r/E_\nu)$, we have:
\begin{equation}
l_{0}^{N} l{_{0}^{N^{\prime }}}^{*} =  c_{0}^{N} c_{0}^{N^{\prime }}\frac{|\vec{q}|^{2}}{m_{N}^{2}} 4E_{\nu }^{2}\left( 1-\frac{E_r}{E_{\nu }}\right) \rho^{2}\,,
\end{equation}
where we use $k^{\prime 0}\approx k^0=E_\nu$. Analogously, other leptonic parts reads:
\begin{equation}
\begin{aligned}
Re\left[ l_{0}^{N} l_{E,3}^{N^{\prime } *}\right] = & {\displaystyle c_{0}^{N} G_{T}^{N^{\prime } ,1}}\frac{|\vec{q}|}{m_{N}} 8E_{\nu }^{2}\left( 1-\frac{E_r}{E_{\nu }}\right) \rho^{2}\\
l_{E,3}^{N} l_{E,3}^{N^{\prime } *} =l_{5,3}^{N} l_{5,3}^{N^{\prime } *} = & {\displaystyle G_{T}^{N,1} G_{T}^{N^{\prime } ,1}} 16E_{\nu }^{2}\left( 1-\frac{E_r}{E_{\nu }}\right) \rho^{2}\\
\frac{1}{2}\left(\vec{l}_{5}^{N} \cdot \vec{l}_{5}^{N^{\prime } *} -l_{5,3}^{N} l_{5,3}^{N^{\prime } *}\right) = & {\displaystyle G_{T}^{N,1} G_{T}^{N^{\prime } ,1}} 8E_{\nu }^{2}\left( 1-\frac{ME_r}{2E_{\nu }^{2}} -\frac{E_r}{E_{\nu }}\right) \rho^{2}\\
\frac{1}{2}\left(\vec{l}_{E}^{N} \cdot \vec{l}_{E}^{N^{\prime } *} -l_{E,3}^{N} l_{E,3}^{N^{\prime } *}\right) = & \frac{1}{2}\left(\vec{l}_{5}^{N} \cdot \vec{l}_{5}^{N^{\prime } *} -l_{5,3}^{N} l_{5,3}^{N^{\prime } *}\right)\,,
\end{aligned}
\end{equation}
}where $l^N_{\alpha,3}$ ($\alpha=5,E$) are the longitudinal component. Note that the longitudinal direction is aligned with $\vec{q}$.
Given $\frac{d\sigma }{dE_r} =\frac{1}{32\pi ME_{\nu }^{2}}\overline{|\mathcal{M} |^{2}}$, we finally obtain Eqn.~\ref{eqn:cross_section}.

\section{\zbl{Two-body currents}}
\label{supp:2b}
\begin{figure}[h]
    \centering
    \includegraphics[width=0.6\linewidth]{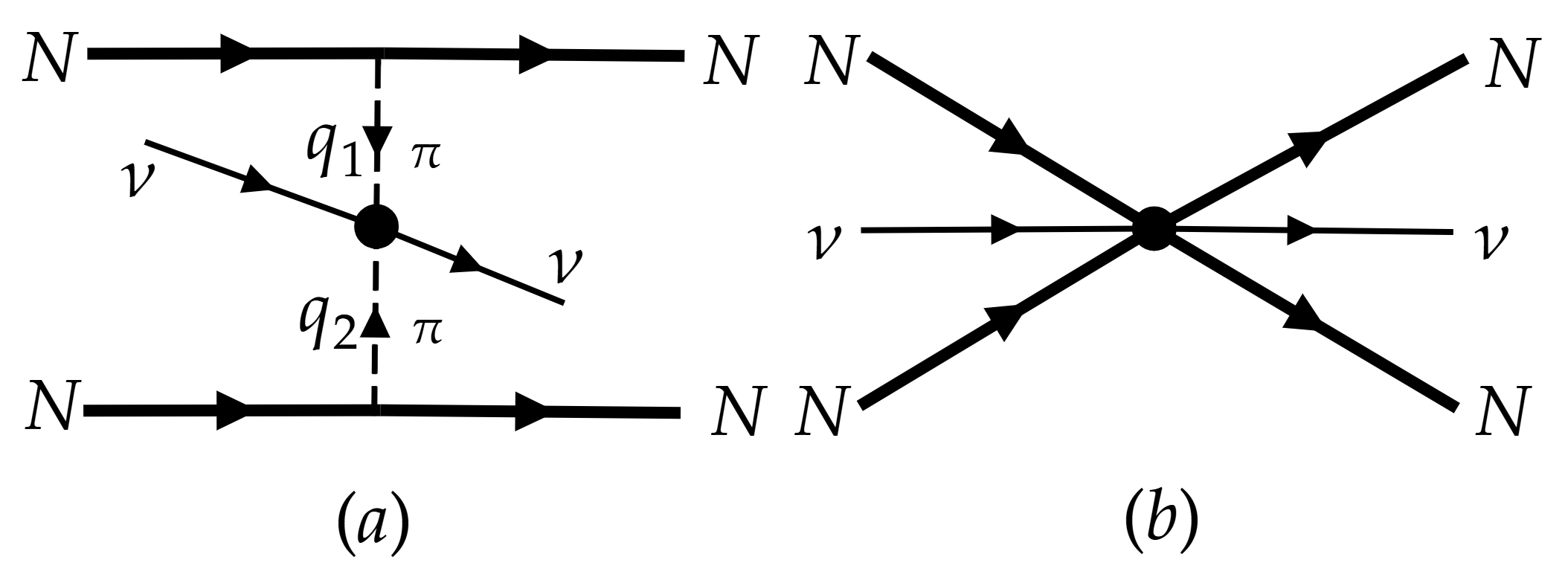}
    \caption{Two-body diagrams that contribute to neutrino-nucleus scattering with tensor interactions. The filled circles represent the tensor interaction vertices.}
    \label{fig:tensor2b}
\end{figure}
The two-body (2b) currents can be derived on the basis of the chiral effective field theory ($\chi$EFT). With the external field method~\cite{Gasser:1983yg, Gasser:1984gg}, we can go from quark couplings to nucleon couplings. To derive a hadronic Lagrangian, we need to write down all terms invariant under the chiral transformation with some proper building blocks. For instance, $U=\exp\left(\frac{i}{F_\pi}\pi^a\tau^a\right)$ is a unitary nonlinear representation of the Goldstone modes for $\mathrm{SU(2)}$ chiral symmetry, where $\pi^a$ are pions with flavor $a$ and $F_\pi$ is the decay constant of the pion in the chiral limit. Following the convention of Refs.~\cite{Ecker:1988te,Cata:2007ns}, we have the following building blocks limited to discussion of tensor interactions:
\begin{equation}
u_{\mu } =iu^{\dagger } D_{\mu } Uu^{\dagger } , T_{+}^{\mu \nu } =u^{\dagger } T^{\mu \nu } u^{\dagger } + uT^{\mu \nu \dagger } u\,,
\end{equation}
where $u=\sqrt{U}$ is the chiral vielbein, and $T^{\mu \nu }$ is the tensor source. In our case, the covariant derivative $D_{\mu }$ can be simplified as $\partial_{\mu }$, because we do not involve vector or axial-vector sources here. The tensor source for Eqn.~\ref{eqn:Lagranian-quark} is $T^{\mu \nu } =\left(\frac{G_{T}^{u} +G_{T}^{d}}{2} 1+\frac{G_{T}^{u} -G_{T}^{d}}{2} \tau ^{3}\right)\left(\overline{\nu } \sigma ^{\mu \nu } \nu \right)$, where the lepton flavor label is omitted. An advantage of these building blocks is that they transform in the same manner under the chiral group, i.e., $hXh^{\dagger}~(X=u_\mu, T_{+}^{\mu \nu })$. Hence, it is convenient to construct the hadronic Lagrangian. 

Given the limited literature on tensor interactions within $\chi$EFT, particularly for 2b currents, we do not provide a comprehensive analysis here. Instead, we focus solely on the lowest-order hadronic Lagrangian~\cite{Cata:2007ns,Filandri:2024azj}:
\begin{equation}
\mathcal{L}_{\pi N}^{\mathnormal{T}} =\tilde{c}_{1}\overline{N} \sigma _{\mu \nu } \langle T_{+}^{\mu \nu } \rangle N+\tilde{c}_{2}\overline{N} \sigma _{\mu \nu }\hat{T}_{+}^{\mu \nu } N-i\tilde{c}_{3} \langle T_{+}^{\mu \nu } u_{\mu } u_{\nu } \rangle \,,
\end{equation}
where $\tilde{c}_i$ are low-energy constants (LECs), $\langle \cdots \rangle$ means the trace of the flavor space, $\hat{T}_{+}^{\mu \nu }\equiv T_{+}^{\mu \nu }-\frac{1}{2}\langle T_{+}^{\mu \nu } \rangle$, and $N$ is a doublet in the isospin space. Note that this chiral Lagrangian is hermitian and invariant under discrete symmetries such as parity and charge conjugation.
The first two terms contribute to the leading one-body current, so their LECs can be determined by the tensor nucleon form factors. We can observe that they are invariant under the transformation $\pi\to -\pi$, resulting in $u\longleftrightarrow u^\dagger$. Therefore, the tensor source have to couple with 2 or more even number of pions, thus the first two terms can not contribute to 2b currents. 
For the $\tilde{c}_3$ term, tensor sources couple with two pions by $-i\frac{\tilde{c}_{3}}{F_{0}^{2}} \epsilon _{abc}\left( T_{+}^{\mu \nu }\right)^{a} \partial _{\mu } \phi ^{b} \partial _{\nu } \phi ^{c}$, as illustrated in Fig.~\ref{fig:tensor2b}~(a). In neutral processes, we only need to consider the isovector part, i.e, $\left( T_{+}^{\mu \nu }\right)^{3}$. Together with the pion-nucleon coupling term in $\chi$EFT, the 2b current of Fig.~\ref{fig:tensor2b}~(a) contributes to $\mathcal{M}_{\mathnormal{2b}}^{\mathnormal{T}} \supset \langle \nu^\prime|\left( T_{+}^{\mu \nu }\right)^{3}|\nu\rangle  J_{\mu \nu }^{( b)}$ and reads:
\begin{equation}
J_{\mu \nu }^{( b)} =-\frac{\tilde{c}_{3} g_{A}^{2}}{4F_{\pi }^{4}}(\vec{\tau }_{1} \times \vec{\tau }_{2})^{3}\left[\frac{\vec{\sigma }_{1} \cdot \vec{q}_{1}}{\vec{q}_{1}^{2} +m_{\pi }^{2}}\frac{\vec{\sigma }_{2} \cdot \vec{q}_{2}}{\vec{q}_{2}^{2} +m_{\pi }^{2}} q_{1\mu } q_{2\nu } -( 1\leftrightarrow 2)\right]\,,
\end{equation}
where $g_A$ is the nucleon’s axial-vector coupling constant, and $m_\pi$ is the pion mass.

In order to estimate the contribution of the 2b current, we convert the 2b current into a normal one-body current by using the independent particle approximation~\cite{Park:1997vv}, i.e., we take the initial (final) wavefunction as the simple product of one-nucleon states $\alpha$ ($\beta$) and the Fock state $|F\rangle$ which describes the Fermi sea: $|F\rangle=\prod_{h\in F}a_{h}^{\dagger}|\mathrm{vac}\rangle$, where $h$ denotes a state in the Fermi sea. The two-body matrix element can be derived by:
\begin{equation}
\langle \beta ;F|J_{\mu \nu } |\alpha ;F\rangle =\sum _{h\in F} [\langle \beta ,h|J_{\mu \nu } |\alpha ,h\rangle -\langle \beta ,h|J_{\mu \nu } |h,\alpha \rangle ]\,,
\end{equation}
where the first term is the direct (or Hartree) term and the second is the exchange (or Fock) term. Moreover, we average over the direction of the probed nucleons and replace the discrete sum with the integral:
\begin{equation}
\langle \beta ;F|J_{\mu \nu } |\alpha ;F\rangle =\sum _{\sigma _{h},\tau_h}\int ^{k_{\mathrm{F}}}\frac{\mathrm{d}^{3} p_h}{(2\pi )^{3}}\int \frac{\mathrm{d} \Omega_{\hat{p}_\alpha}}{4\pi } [\langle \beta ,h|J_{\mu \nu } |\alpha ,h\rangle -\langle \beta ,h|J_{\mu \nu } |h,\alpha \rangle ]\,.
\end{equation}
The one-nucleon state $\alpha$ is on the Fermi surface: $|\vec{p}_\alpha|=k_\mathrm{F}$ ($k_\mathrm{F}$ is the Fermi momentum), while $\beta$ is scattered: $\vec{p}_\beta=\vec{p}_\alpha-\vec{q}$. In the Hartree term, momentum transfers are $\vec{q_1}=-\vec{q}$ and $\vec{q_2}=0$, since the nucleons in the occupied state $\vec{p}_h$ are identical. Therefore, Fig.~\ref{fig:tensor2b}~(a) only contributes to the Fock term, where $\vec{q_1}=\vec{p}_h-\vec{p}_\alpha$ and $\vec{q_2}=\vec{p}_\beta-\vec{p}_h=\vec{p}_\alpha-\vec{q}-\vec{p}_h$. Focusing on the unit operator, the matrix elements read:
\begin{equation}
\langle \beta ;F|J_{\mu \nu } |\alpha ;F\rangle \simeq -\frac{i\tilde{c}_{3} g_{A}^{2}}{2F_{\pi }^{4} m_{\pi }^{4}}\int ^{k_{\mathrm{F}}}\frac{\mathrm{d}^{3} p_{h}}{(2\pi )^{3}}\int \frac{\mathrm{d} \Omega_{\hat{p}_{\alpha }}}{4\pi } \langle \beta |\tau ^{3} 1_{N} |\alpha \rangle [\vec{q}_{1} \cdot \vec{q}_{2} q_{1\mu } q_{2\nu } -( 1\leftrightarrow 2)]\,,
\end{equation}
where we assume $m_\pi^2$ dominates in the propagators, and we use the following relation:
\begin{equation}
\begin{aligned}
\sum _{\sigma _{h} ,\tau _{h}} \langle \beta ,h|(\vec{\tau }_{1} \times \vec{\tau }_{2})^{3}\vec{\sigma }_{1} \cdot \vec{q}_{1}\vec{\sigma }_{2} \cdot \vec{q}_{2} |h,\alpha \rangle  & =\sum _{\sigma _{h} ,\tau _{h}} \varepsilon ^{3ab} \langle \beta |\tau ^{b}\vec{\sigma } \cdot \vec{q}_{2} |h\rangle \langle h|\tau ^{a}\vec{\sigma } \cdot \vec{q}_{1} |\alpha \rangle \\
 & =-2i\varepsilon ^{3ab} \langle \beta |\tau ^{3}(\vec{q}_{1} \cdot \vec{q}_2 -i\vec{\sigma } \cdot (\vec{q}_{1} \times \vec{q}_{2}) |\alpha \rangle \,.
\end{aligned}
\end{equation}
Since we require $J_{\mu\nu}$ to be anti-symmetric, $q_{1\mu}$ and $q_{2\nu}$ must have different indexes. Once the directions of $\vec{p}_\alpha$ and $\vec{p}_h$ are averaged, we obtain the null result. 
This outcome is further corroborated by examining the zero momentum transfer limit ($\vec{q}=0$). In this scenario, $\vec{q_1}=-\vec{q}_2$, leading to a vanishing contribution from the 2b current due to its inherent anti-symmetric property.

Except for the lowest order chiral Lagrangian, one can also consider the contact term:
\begin{equation}
\mathcal{L}_{\mathnormal{contact}}^{\mathnormal{T}} =\tilde{d}_{1}\overline{N} \sigma _{\mu \nu } \langle T_{+}^{\mu \nu } \rangle N\overline{N} N+\tilde{d}_{2}\overline{N} \sigma _{\mu \nu }\hat{T}_{+}^{\mu \nu } N\overline{N} N\,,
\end{equation}
where $\tilde{d}_i$ are new LECs. The corresponding diagram is depicted in Fig.~\ref{fig:tensor2b}~(b), and its associated 2b currents that contribute to $\mathcal{M}_{\mathrm{2b}}^{\mathrm{T}} \supset 2\langle \nu ^{\prime } |\left( T_{+}^{\mu \nu }\right)^{0} |\nu \rangle J_{\mu \nu }^{(c1)} +\langle \nu ^{\prime } |\left( T_{+}^{\mu \nu }\right)^{3} |\nu \rangle J_{\mu \nu }^{(c2)}$ are:
\begin{equation}
J_{0i}^{(c1)} =-\frac{i\tilde{d}_{1}\left( q_{1}^{i} +q_{2}^{i}\right)}{2m_{N}} , J_{0i}^{(c2)} =-\frac{i\tilde{d}_{2}\left( q_{1}^{i} \tau _{1}^{3} +q_{2}^{i} \tau _{2}^{3}\right)}{2m_{N}}\,,
\end{equation}
where we only consider the contributions relevant to the unit operator.
Following the procedure dealing with Fig.~\ref{fig:tensor2b}~(a), we observe that the Fock term vanishes if the direction of $\vec{p}_\alpha$ and $\vec{p}_h$ are averaged over, since the contribution linearly depends on $\vec{q_1}=\vec{p}_h-\vec{p}_\alpha$ and $\vec{q_2}=\vec{p}_\alpha-\vec{q}-\vec{p}_h$. Therefore, only the Hartree term remains:
\begin{equation}
\langle \beta ;F|J_{0i}^{(c1)} |\alpha ;F\rangle =-\frac{i\tilde{d}_{1} \rho q^{i}}{2m_{N}} \langle \beta |1_{N} |\alpha \rangle , \langle \beta ;F|J_{0i}^{(c2)} |\alpha ;F\rangle =-\frac{i\tilde{d}_{2} \rho q^{i}}{2m_{N}} \langle \beta |\tau ^{3} 1_{N} |\alpha \rangle \,,
\end{equation}
where $\rho=2k_{\mathrm{F}}^{3}/3\pi^{2}$ is the density of symmetric nuclear matter. Therefore, $c_0^N$ in Eqn.~\ref{eqn:cross-section-odd} should be modified:
\begin{equation}
c_{0}^{p}\rightarrow c_{0}^{p} +\left[ G_{T}^{u}\left( 2\tilde{d}_{1} +\tilde{d}_{2}\right) +G_{T}^{d}\left( 2\tilde{d}_{1} -\tilde{d}_{2}\right)\right] \rho , c_{0}^{n}\rightarrow c_{0}^{n} +\left[ G_{T}^{u}\left( 2\tilde{d}_{1} -\tilde{d}_{2}\right) +G_{T}^{d}\left( 2\tilde{d}_{1} +\tilde{d}_{2}\right)\right] \rho  \,.
\end{equation}
For~normal~nuclear~matter, $k_{\mathrm{F}}\simeq1.36~\mathrm{fm}^{-1}$ and $\rho\simeq0.17~\mathrm{fm}^{-3}\approx1.3\times10^{-3}~\mathrm{GeV}^3$. Thus, a substantial contribution from the 2b currents requires the two LECs $\tilde{d}_{1}$ and $\tilde{d}_{2}$ to be of the order $\mathcal{O}(10^3)~\mathrm{GeV}^{-3}$.

In conclusion, we estimate the potential contributions of 2b currents based on $\chi$EFT and the independent particle approximation~\cite{Park:1997vv}. Since LECs ( $\tilde{d}_1$,$\tilde{d}_2$) are not determined by experimental data or lattice QCD, we do not include this analysis in our main text. It is shown that configurations depicted in Fig.~\ref{fig:tensor2b}~(b) could potentially yield sizable contributions once the aforementioned LECs are large enough. Moreover, our calculation on 2b currents can be further improved using a naive shell model method instead of the independent particle approximation~\cite{Hoferichter:2016nvd,Hoferichter:2018acd}.

\section{Cross check with spin-suppressed result}
\label{supp:check}
Here, we present how to recover the cross section in the spin-suppressed result~\cite{Hoferichter:2020osn} with Eqn.~\ref{eqn:cross-section-even}. According to the convention in Ref.~\cite{Hoferichter:2020osn}, we define four column matrices: $g^{01} =\begin{pmatrix}
g_{T,1}^{0} & g_{T,1}^{1}
\end{pmatrix}^{T}$, $g^{pn} =\begin{pmatrix}
g_{T,1}^{p} & g_{T,1}^{n}
\end{pmatrix}^{T}$, $\mathcal{F}^{\Sigma ^{\prime } ,\pm } =\begin{pmatrix}
\mathcal{F}_{+}^{\Sigma ^{\prime }} & \mathcal{F}_{-}^{\Sigma ^{\prime }}
\end{pmatrix}^{T}$, and $\mathcal{F}^{\Sigma ^{\prime } ,pn} =\begin{pmatrix}
\mathcal{F}_{p}^{\Sigma ^{\prime }} & \mathcal{F}_{n}^{\Sigma ^{\prime }}
\end{pmatrix}^{T}$. Note that in this convention, $ F_{\Sigma \prime}^{N}(0) =\sqrt{\frac{2}{3}}\sqrt{\frac{( 2J_i+1)( J_i+1)}{4\pi J_i}} \langle S_{N} \rangle$. The transition from the nucleon basis to the isospin basis can then be expressed readily: $g_{i}^{01} =\sum _{j}\mathbf{O}_{ij} g_{j}^{pn}$ and $\mathcal{F}_{i}^{\Sigma ^{\prime } ,\pm } =\sum _{j}\mathbf{O}_{ij}^{-1}\mathcal{F}_{j}^{\Sigma ^{\prime } ,pn}$, with $\mathbf{O} =\frac{1}{2}\begin{pmatrix}
1 & 1\\
1 & -1
\end{pmatrix} =\frac{1}{2}\mathbf{O}^{-1} =\mathbf{O}^{T}$, leading to a useful relation: $\sum _{i}  g_{i}^{pn}{\displaystyle \mathcal{F}_{i}^{\Sigma ^{\prime } ,pn} =}\sum _{i} g_{i}^{01}\mathcal{F}_{i}^{\Sigma ^{\prime } ,\pm }$.
In our conventions, $G_{T,1}^{N}=g_{T,1}^{N}$, and $F_{\Sigma \prime }^{N}(0) =\sqrt{\frac{8( J_i+1)}{3J_i}} \langle S_{N} \rangle$~\cite{DelNobile:2021wmp}. Thus a replacement  $F^{\Sigma ^{\prime }}_{N}\rightarrow F^{\Sigma ^{\prime }}_{N}\sqrt{ \frac{16\pi}{2J_i+1}}$ should be included when converting our result into others. 
For the term proportional to $\mathcal{F}_{N}^{\Sigma ^{\prime }}\mathcal{F}_{N^{\prime }}^{\Sigma ^{\prime }}$, we have:
\begin{equation}
\begin{aligned}
\frac{M}{\pi }\sum _{i,j}  g_{i}^{pn}{\displaystyle \mathcal{F}_{i}^{\Sigma ^{\prime } ,pn} g_{j}^{pn}\mathcal{F}_{j}^{\Sigma ^{\prime } ,pn}\left( 1-\frac{ME_r}{2E_{\nu }^{2}} -\frac{E_r}{E_{\nu }}\right)} & =\frac{16M}{2J_i+1}\sum _{i,j}  {\displaystyle g_{i}^{01}\mathcal{F}_{i}^{\Sigma ^{\prime } ,\pm } g_{j}^{01}\mathcal{F}_{j}^{\Sigma ^{\prime } ,\pm }\left( 1-\frac{ME_r}{2E_{\nu }^{2}} -\frac{E_r}{E_{\nu }}\right)}\\
 & =\frac{16M}{2J_i+1}{\displaystyle \left( 1-\frac{ME_r}{2E_{\nu }^{2}} -\frac{E_r}{E_{\nu }}\right)}\left[ (g_{T,1}^{0} )^{2}\overline{S}_{00}^{\mathcal{T}} +g_{T,1}^{0} g_{T,1}^{1}\overline{S}_{01}^{\mathcal{T}} +(g_{T,1}^{1} )^{2}\overline{S}_{11}^{\mathcal{T}}\right]\,,
\end{aligned}
\end{equation}
where the left hand size is from Eqn.~\ref{eqn:cross-section-even}, and we use the relation: $\overline{S}_{00}^{\mathcal{T}} =\left(\mathcal{F}_{+}^{\Sigma ^{\prime }}\right)^{2} $, $ \overline{S}_{11}^{\mathcal{T}} =\left(\mathcal{F}_{-}^{\Sigma ^{\prime }}\right)^{2} $ and $ \overline{S}_{01}^{\mathcal{T}} =2\mathcal{F}_{+}^{\Sigma ^{\prime }}\mathcal{F}_{-}^{\Sigma ^{\prime }}$. This yields the same transverse contribution of  Eqn.~108 in Ref.~\cite{Hoferichter:2020osn}.
The longitudinal one can be done in the same way:
\begin{equation}
\frac{M}{\pi }\sum _{i,j}  g_{i}^{pn}{\displaystyle \mathcal{F}_{i}^{\Sigma ^{\prime ^{\prime }} ,pn} g_{j}^{pn}\mathcal{F}_{j}^{\Sigma ^{\prime ^{\prime }} ,pn} 2\left( 1-\frac{E_r}{E_{\nu }}\right)} =\frac{32M}{2J_i+1}{\displaystyle \left( 1-\frac{E_r}{E_{\nu }}\right)}\left[ (g_{T,1}^{0} )^{2}\overline{S}_{00}^{\mathcal{L}} +g_{T,1}^{0} g_{T,1}^{1}\overline{S}_{01}^{\mathcal{L}} +(g_{T,1}^{1} )^{2}\overline{S}_{11}^{\mathcal{L}}\right]\,.
\end{equation}

\section{Experimental analysis}
\label{supp:analysis}
The total differential cross section for isotope $i$ reads:
\begin{equation}
\frac{d\sigma _{i}}{dE_r} =\frac{d\sigma _{i,\mathrm{SM}}}{dE_r} +\frac{d\sigma _{i,T}}{dE_r}\,,
\end{equation}
where $\frac{d\sigma _{i,T}}{dE_r}$ presented in Eqn.~\ref{eqn:cross_section}, and $\frac{d\sigma _{i,\mathrm{SM}}}{dE_r} $ is taken from Ref.~\cite{Hoferichter:2020osn}: 
\begin{equation}
\frac{\mathrm{d} \sigma _{i,\mathrm{SM}}}{\mathrm{d} E_r} =\frac{G_{F}^{2} M}{4\pi }\left( 1-\frac{ME_r}{2E_{\nu }^{2}}\right) Q_{\mathrm{w}}^{2} F_{\mathrm{w}} (|\vec{q} |^{2} )^{2}\,,
\end{equation}
where $Q_\mathrm{w}=Z(1-4\mathrm{sin}^2\theta_W)-N$ is the weak charge, and $F_{\mathrm{w}}$ is the weak charge distribution inside the nucleus~\cite{Hoferichter:2020osn}. Note that the form factor parametrization is inconsequential for current CE$\nu$NS experiments, yet it will become relevant for future high-precision CE$\nu$NS experiments~\cite{AristizabalSierra:2019zmy}. \zbl{It is also worthy to note that the weak form factor can be derived by different methods, including couple cluster theory~\cite{Payne:2019wvy},  relativistic mean-field model~\cite{Yang:2019pbx}, solving Hartree-Fock equations~\cite{VanDessel:2020epd,Co:2020gwl}, microscopic picture~\cite{Tomalak:2020zfh}, etc.}

\begin{figure}
    \centering
    \includegraphics[width=0.4\linewidth]{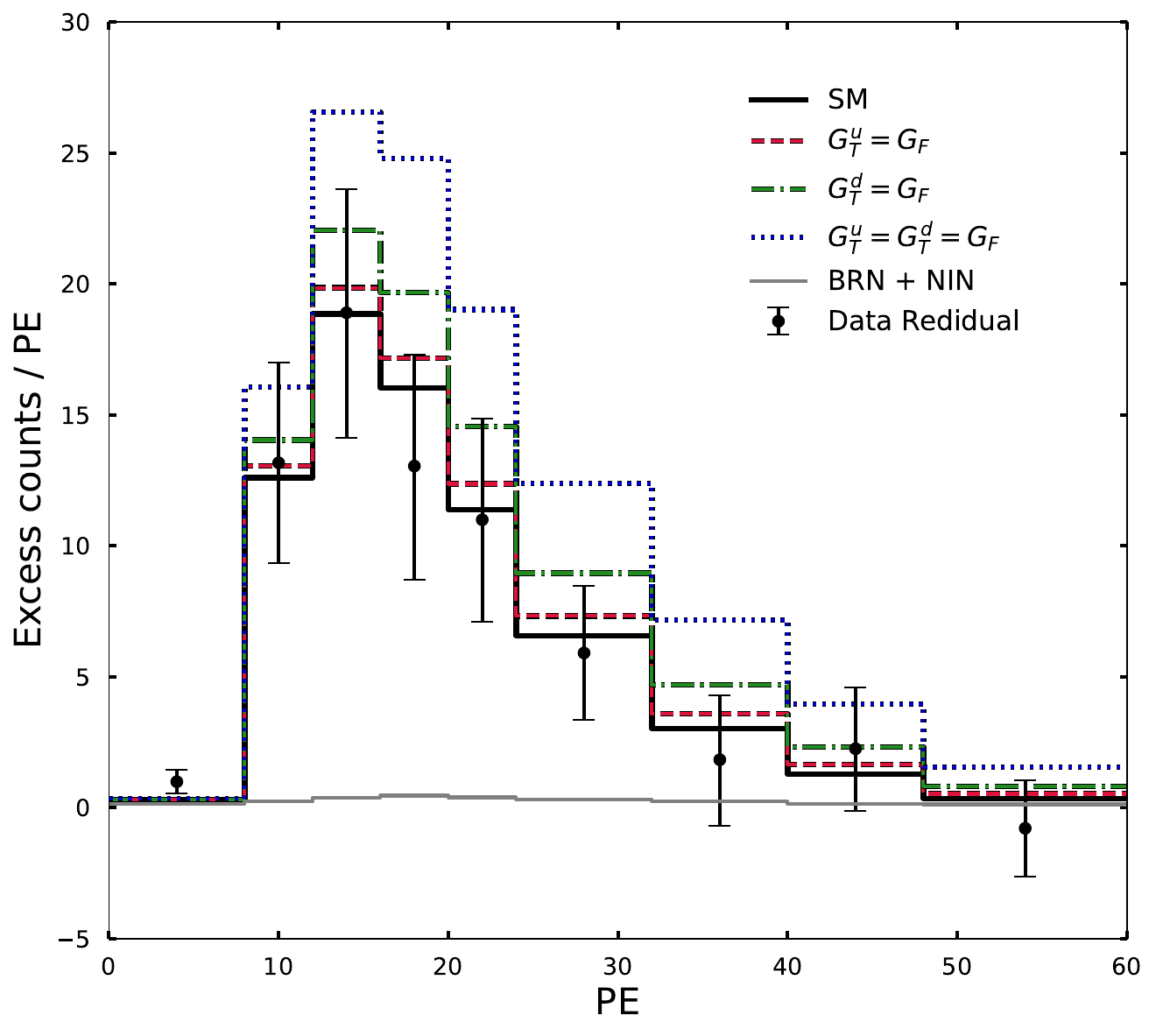}
    \caption{COHERENT-2021 data and theoretical predicted spectra.}
    \label{fig:spectrum}
\end{figure}

In COHERENT, the low-energy neutrino fluxes are generated at the Spallation Neutron Source (SNS) at the Oak Ridge National Laboratory. 
The neutrino fluxes consist of monochromatic $\nu_\mu$ coming from $\pi^+$ decays, $\pi^+\to\mu^+\nu_\mu$, along with delayed $\nu_e$ and $\bar{\nu}_\mu$ from the subsequent $\mu^+$ decays, $\mu^+\to e^+\bar{\nu}_\mu\nu_e$. 
Convoluting the differential cross section with neutrino fluxes $\Phi _{\alpha }$, the differential rates for the CsI detector reads:
\begin{equation}
\frac{dR_{\mathrm{CsI}}}{dE_r} =\sum _{i=\mathrm{Cs,I}}\sum _{\alpha }\frac{1}{M_{i}}\int _{E_{\nu }^{\min}}^{E_{\nu }^{\max}}\frac{d\Phi _{\alpha }}{dE_{\nu }}\frac{d\sigma _{i}}{dE_r} dE_{\nu }\,,
\end{equation}
where $E_{\nu }^{\mathrm{min}}\simeq \sqrt {ME_r/ 2}$, $E_{\nu}^{\mathrm{max}}=m_\mu/2=52.8~\mathrm{MeV}$, $M_i$ is the isotope mass, and summations are over $\alpha = \nu_\mu, \nu_e, \bar{\nu}_\mu$.
To accurately predict the total events in the CsI detector as a function of the reconstructed energy $E_\mathrm{rec}$ in units of PE, quenching, energy smearing and detection efficiency must be taken into account. A empirically parameterized quenching $E_{ee}=E_{ee}(E_r)$ builds the relationship between the nuclear recoil energy $T$ and the true quenched energy deposition $E_{ee}$. 
Considering the energy smearing $P(E_\mathrm{rec},E_{ee})$ and the efficiency $\varepsilon_{E}(E_\mathrm{rec})$, we have the differential recoil spectrum as a function of $E_\mathrm{rec}$:
\begin{equation}\label{eqn:rate-CsI}
\frac{dR}{dE_{\mathrm{rec}}} =\int dE_{ee} P(E_{\mathrm{rec}} ,E_{ee} )\frac{dE_r}{dE_{ee}}\frac{dR_{\mathrm{CsI}}}{dE_r}\,.
\end{equation}
In addition, because the reconstructed energy and time are uncorrelated, we simply convolute the time information with the time efficiency given by Ref.~\cite{COHERENT:2021xmm}, leading to $\epsilon^t_{\nu_\mu}=0.997$, $\epsilon^t_{\bar{\nu}_\mu}=0.848$ and $\epsilon^t_{\nu_e}=0.848$.
Neutrino fluxes, quenching, energy smearing and detection efficiency can be found in the supplementary material of Ref.~\cite{COHERENT:2021xmm}.
For the COHERENT-2021 data, we adopt the following $\chi^2$:
\begin{equation}\label{eqn:chi2-COHERENT}
\chi _{\mathrm{COHE.}}^{2} =\underset{\alpha , \beta }{\min}\left\{\sum _{i=2}^{8}\left[\frac{N_{\mathrm{meas}}^{i} -(1+\alpha )N_{\mathrm{th}}^{i} -(1+\beta )B_{\mathrm{on}}^{i}}{\sigma _{\mathrm{stat}}^{i}}\right]^{2} +\left(\frac{\alpha }{\sigma _{\alpha }}\right)^{2} +\left(\frac{\beta }{\sigma _{\beta }}\right)^{2}\right\}\,,
\end{equation}
where $N_{\mathrm{meas}}^{i}$ is the measured number of events per energy bin, $B_{\mathrm{on}}^i$ is the beam-on background, $N_{\mathrm{th}}^i$ is the theoretically predicted number of events per energy bin in the presence of tensor interactions, $\sigma_{\mathrm{stat}}^i$ is the statistical uncertainty per energy bin, $\sigma_\alpha=0.1195$ is the total systematic uncertainty, and $\sigma_\beta=0.25$ is the systematic uncertainty for the background. \zbl{Furthermore, discrepancies emerge in the nuclear response functions when employing various nuclear shell model interactions at high recoil energies~\cite{AbdelKhaleq:2023ipt, AbdelKhaleq:2024hir}. Given that the dominant contribution stems from $\mathcal{F}_{N}^{M}$, we assume the  theoretical uncertainty $\sigma_\mathrm{th}\simeq0.1$, yielding  $\sigma_\alpha\to\sqrt{\sigma_\alpha^2+\sigma_\mathrm{th}^2}\approx 0.156$. } $N_{\mathrm{th}}^i$ are derived by integrating Eqn.~\ref{eqn:rate-CsI} over the energy bin $i$. We present the data and theoretical spectra in Fig.~\ref{fig:spectrum}.

\begin{table}
    \centering
    \begin{tabular}{|c|c|c|c|} \hline 
         Dataset&  Energy range [keV]&  $N_\mathrm{meas}$& $N_\mathrm{th}$\\ \hline 
         PandaX-4T (paired)&  [1.1, 3.0]&  $3.5\pm 1.3$& 2.4\\ \hline 
         PandaX-4T (US2)&  [0.3, 3.0]&  $75\pm 28$& 44.3\\ \hline 
         XENONnT&  [0.5, 3.0]&  $10.7\pm 3.95$& 11.2\\ \hline
    \end{tabular}
    \caption{Measurements on $^8$B from PandaX (paired and US2) and XENON experiments.}
    \label{tab:B8-data}
\end{table}

The differential event rates for a Xenon-based detector reads:
\begin{equation}\label{eqn:rates-B8}
\frac{dR_{\mathrm{Xe}}}{dE_r} =\sum _{i}\sum _{\alpha =e,\mu ,\tau }\frac{1}{M_{i}}\int _{E_{\nu }^{\min}}^{E_{\nu }^{\max}}\left[\int ^{R_{\odot }}\frac{d^{2} \Phi _{^{8} \mathrm{B}}}{dE_{\nu } dr} P_{e\alpha }( r) dr\right]\frac{d\sigma _{i,\nu _{\alpha }}}{dE_r} dE_{\nu }\,,
\end{equation}
where $P_{e\alpha}$ denotes  the neutrino survival probability from the Sun to the Earth, $r$ is the solar neutrino production position, $R_{\odot }$ represents the solar radius, and the summation spans naturally occurring xenon isotopes in the xenon-based detector, weighted by their terrestrial abundances: $^{128}$Xe (1.92\%), $^{129}$Xe (26.44\%), $^{130}$Xe (4.08\%), $^{131}$Xe (21.18\%), $^{132}$Xe (26.89\%), $^{134}$Xe (10.44\%), and $^{136}$Xe (8.87\%). 
The $^8$B neutrino flux $\frac{d^{2} \Phi _{^{8} \mathrm{B}}}{dE_{\nu } dr}$ is taken from BS05 standard solar model~\cite{Bahcall:2004pz},  with a normalization factor of $5.16\times 10^{6}~\mathrm{cm^{-2}s^{-1}}$ from the global fits of neutrino data~\cite{Bergstrom:2016cbh}. 
In the scenario of the $\tau$-only tensor interaction, the matter effect in the neutrino propagation has been taken into account. Since the tensor interactions only affect the neutrino propagation in a polarized medium~\cite{Bergmann:1999rz}, the matter effect is exclusively determined by the SM weak interaction. For non-standard neutrino vector interactions, their influence on the neutrino propagation should be incorporated~\cite{AristizabalSierra:2024nwf,Li:2024iij}. Therefore, the neutrino survival probability can be evaluated by:
\begin{equation}
P_{\alpha \beta }(r) =\sum _{k} |U_{\alpha k}^{\mathrm{mat.}}(r) |^{2} |U_{\beta k}^{\mathrm{PMNS}} |^{2}\,,
\end{equation}
where $U^{\mathrm{PMNS}}$ is the PMNS matrix and $U^{\mathrm{mat.}}$ can be generated by diagonalize the SM matter effect Hamiltonian. 
One more step to derive the theoretical predicted $^8$B event numbers is to convolute Eqn.~\ref{eqn:rates-B8} with the efficiencies from PandaX and XENONnT experiments~\cite{PandaX:2024muv, XENON:2024ijk}. 
Three measurements are listed in Tab.~\ref{tab:B8-data}, as well as our SM predictions. The $\chi^2$ for $^8$B observations is defined by:
\begin{equation}\label{eqn:chi2-b8}
\chi _{^{8}\mathrm{B}}^{2} =\underset{\alpha }{\min}\left\{\sum _{j}\left[\frac{N_{\mathrm{meas}}^{j} -(1+\alpha )N_{\mathrm{th}}^{j}}{\sigma _{\mathrm{stat}}^{j}}\right]^{2} +\left(\frac{\alpha }{\sigma _{\alpha }}\right)^{2}\right\}\,,
\end{equation}
where $j$ labels the dataset in Tab.~\ref{tab:B8-data}.
Analogous to COHERENT-2021 data, we set $\sigma_\alpha=0.12$ to account for the systematic uncertainty, originating from the prediction of the neutrino flux~\cite{Vitagliano:2019yzm}. According to Tab.~\ref{tab:B8-data}, the statistical uncertainties for three datasets are $37.1\%$, $37.3\%$ and $36.9\%$, respectively.

\end{document}